\documentclass[useAMS,usenatbib,usegraphicx,twocolumn]{mn2e}

\title[Structure of dynamical condensation fronts
]{Structure of dynamical condensation fronts in the interstellar medium 
}

\author[Kazunari Iwasaki and Shu-ichiro Inutsuka]{Kazunari Iwasaki$^{1}$\thanks{E-mail:
iwasaki@nagoya-u.jp, inutsuka@nagoya-u.jp} and Shu-ichiro Inutsuka$^{1}$\footnotemark[1]
\\
$^{1}$Department of Physics, Nagoya University, Furo-cho, 
Chikusa-ku, Nagoya, Aichi, 464-8602, Japan}
\begin{document}

\date{Accepted 2012 April 22. Received 2012 April 22; in original form 2012 March 16}

\pagerange{\pageref{firstpage}--\pageref{lastpage}} \pubyear{2002}

\maketitle

\label{firstpage}

\begin{abstract}
     In this paper, we investigate the structure of condensation fronts
     from warm diffuse gas to cold neutral medium (CNM)
     under the plane parallel geometry. 
     The solutions have two parameters, the pressure of the CNM
     and the mass flux across the transition front, and their ranges are much wider 
     than previously thought.
     First, we consider the pressure range where the three phases, the CNM, the unstable phase, and the warm 
     neutral medium, can coexist in the pressure equilibrium.
     In a wide range of the mass flux, we find solutions connecting 
     the CNM and the unstable phase. 
     Moreover, we find solutions in larger pressure range where there is only one 
     thermal equilibrium state or the CNM.
     These solutions can be realized in shock-compressed regions that are promising sites
     of molecular cloud formation. We also find remarkable 
     properties in our solutions. 
     Heat conduction becomes less important with increasing mass flux, and the 
     thickness of the transition layer 
     is characterized by the cooling length instead of the Field length.
\end{abstract}

\begin{keywords}
hydrodynamics - ISM: clouds - ISM: kinematics and dynamics
\end{keywords}

\section{Introduction}
It is well known that in the interstellar medium (ISM), 
a clumpy low-temperature phase [cold neutral medium (CNM)] and a diffuse 
high-temperature phase [warm neutral medium (WNM)]
can coexist 
in pressure equilibrium as a result of the balance of radiative cooling and 
heating owing to external radiation fields and cosmic rays 
\citep{Fetal69,Wetal95,Wetal03}.
The CNM of atomic gas is observed as HI cloud 
($n\sim10-100$ cm$^{-3}$, $T\sim10^2$ K).  
These two phases are thermally stable. On the other hand, 
the thermal instability (TI) arises in 
the temperature range between these two phases.
Thus, the ISM in atomic phase can be interpreted as bistable fluid. 

Pioneering work on the TI has been done by \citet{F65} who performed a linear analysis of 
the thermal equilibrium gas and derived a simple criterion for the TI.
Focusing on a fluid element, \citet{B86} derived a criterion for the TI of thermal non-equilibrium gas.
In the nonlinear evolution, \citet{IT08} found a family of self-similar solutions describing the 
condensation of radiative gas layer assuming a simple power-law cooling rate.
\citet{IT09} investigated the linear stability of the self-similar solutions, 
and they suggested that 
the condensing layer will fragment in various scales as long as the transverse 
scale is larger than the Field length.
Physics of the bistable fluid has been investigated by many authors.
\citet[][hereafter ZP69]{ZP69} and \citet{PB70} investigated the structure of the transition front 
connecting the CNM and WNM under the plane-parallel geometry.
The thickness of the transition front is characterized by the Field length below which 
the TI is stabilized by heat conduction \citep{F65}.
They found that a static solution is obtained at a so-called saturation 
pressure. The transition front becomes a condensation (evaporation) front 
if surrounding pressure is larger (less) than the saturation pressure.
These properties of the transition front depend on its geometry.
In the spherical symmetric geometry,
\citet{GL73} investigated isobaric flows. They found that spherical clouds have a minimum size below 
which they inevitably evaporate \citep[more general description is found in][]{Netal05}.
Linear stability of the transition layers has been investigated by \citet{IIK06} in the plane-parallel geometry, 
and they found that the evaporation front is unstable against 
corrugation type fluctuation, while the condensation 
front is stable. 
\citet{SZ09} have considered the effect of magnetic field on the instability.
The physical mechanism of the corrugation instability is analogous to
the Darriues-Landau instability \citep{LL87}.
Taking into account magnetic field perpendicular to the normal of the front, 
\citet{SZ10} investigated linear stability of transition layers.

In the above-mentioned theoretical works with respect to the bistable fluid, the phase transition proceeds
in a quasi-static manner through the heat conduction. 
However, recent multidimensional numerical simulations show more 
a dynamical turbulent structure.  \citet{KN02} have investigated 
the non-linear development of the TI in the three-dimensional hydrodynamical simulations with 
periodic boundary condition without any external forcing. As the initial condition, they set
the hot ionized medium ($T=2\times 10^6$ K) where cooling dominates heating.
The gas is quickly decomposed into two stable phases (CNM and WNM) and 
the intermediate unstable phase. In the multiphase medium,
the supersonic turbulence is developed by conversion from the thermal energy to the 
kinetic energy. 
They found that turbulence decays on a dynamical timescale. This decaying timescale is larger 
than that in supersonic isothermal turbulence of one-phase medium where 
it is smaller or comparable to the flow crossing time 
\citep{Setal98}.
\citet{KI06} have done similar calculations on a much longer timescale.
As the initial condition, they set an unstable gas in thermal equilibrium state with 
density fluctuations. 
They found a self-sustained turbulence with velocity dispersion of $\sim0.2-0.4$ km s$^{-1}$ 
for a period of at least their simulation time (several 10 Myr)
in the bistable fluid after temporal decaying found in \citet{KN02}.
External forcing such as shock compression can drive stronger long-lived turbulence. 
\citet{KI02} investigated development of the TI in a shock compressed region by using a two-dimensional
hydrodynamical simulation. They found that the velocity dispersion is as large as 
several km s$^{-1}$,  which is larger than the sound speed of the CNM.
The supersonic turbulent motion is maintained as long as the shock wave 
continutes to propagate and provide postshock gas.
They suggested that this supersonic translational motion of the CNM 
is observed as the supersonic turbulence in the ISM.
The cloudlets are precursor of molecular clouds
because they are as dense as molecular clouds.  After their work,
many authors have investigated the formation of the CNM 
in shocked gases in two-dimensional calculations \citep{AH05,HA07,Hetal05,Hetal06}, and 
in three-dimensional calculations \citep{Getal05,Vetal06,AH10}.
The influence of the magnetic field on the development of the TI has 
been investigated by \citet{II08,II09}, \citet{Hene08}, \citet{Hetal09}.

Although physics of the bistable fluid is developing, the physical mechanism of driving turbulence 
is not fully understood.
As mentioned above, recent numerical simulations have found that the turbulence involving the TI
is dynamical compared with the ZP69 picture that predicts very 
slow transition between CNM and WNM through the heat conduction.
Moreover, in the shock compressed region, because of its high pressure, 
the bistable fluid cannot exist. The ZP69 picture cannot be applied directly 
to the TI in the shock compressed region.

In this paper, as a first step to understand the dynamical turbulent structure, 
we develop the works of ZP69 into a more dynamical transition front.
In the steady solutions, there are two parameters, the pressure of the CNM and the mass flux
across the front. ZP69 considered solutions only on a line in the parameter space. 
\citet{Eetal92} found steady solutions with various mass flux. However, 
they used a simple cubic function of the cooling rate that enables them to investigate analytically,
and they assumed spatially constant pressure.
We systematically search steady solutions in larger parameter space 
even in the high pressure range where the bistable fluid does not exist by using a realistic cooling rate. 

This paper is organized as follows: Basic equations and numerical methods are described in section
\ref{sec:numerical}. In section \ref{sec:ZP}, we briefly review the solutions connecting 
the CNM and WNM derived by ZP69.
In section \ref{sec:dynamical}, we present solutions describing 
dynamical transition layer that has larger mass flux than ZP69.
The results are discussed in section \ref{sec:discuss} and are summarized in section \ref{sec:summary}.

\section{Basic Equations and Numerical Methods}\label{sec:numerical}
\begin{figure}
        \begin{center}
                  \includegraphics[width=8.0cm]{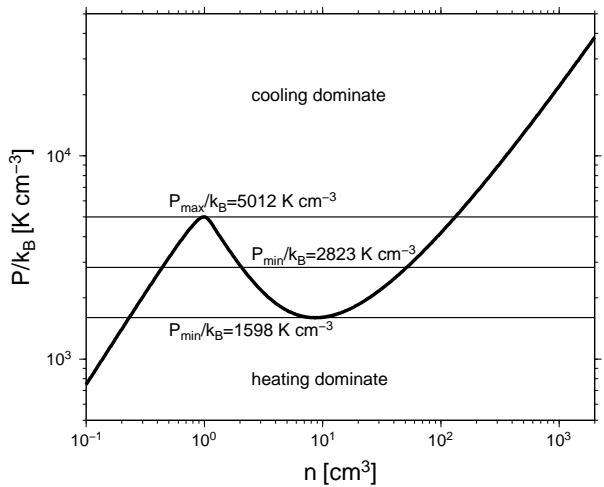}
        \end{center}
        \caption{
        Thermal equilibrium state of the net cooling rate ${\cal L}(\rho,T)$ in $(P,n)$ plane.
        }
        \label{fig:eq}
\end{figure}

Basic equations for radiative gas under the plane-parallel geometry are the continuity equation,
\begin{equation}
        \frac{\partial \rho}{\partial t} + \frac{\partial}{\partial x}(\rho v)=0,
        \label{eoc}
\end{equation}
the momentum conservation,
\begin{equation}
        \frac{\partial \rho v}{\partial t} + \frac{\partial}{\partial x}\left( P + \rho v^2 \right)=0,
        \label{eom}
\end{equation}
the energy equation,
\begin{equation}
        \frac{\partial E}{\partial t} +
        \frac{\partial}{\partial x}\left[ \left( E+P \right)v - \kappa \frac{\partial T}{\partial x} \right]
        = - \rho{\cal L}(\rho,T),
        \label{eoe}
\end{equation}
and the equation of state,
\begin{equation}
        P={\cal R} \rho T
        \label{eos}
\end{equation}
where $E=\rho v^2/2 + P/(\gamma-1)$ is the total energy, and $\gamma=5/3$ is the ratio of specific heats, 
$\kappa$ is the coefficient of heat conductivity, and ${\cal L}(\rho,T)$ is the net cooling rate per unit mass, 
${\cal R}=k_\mathrm{B}/m_\mathrm{H}$ is the gas constant, and 
$m_\mathrm{H}$ is the hydrogen mass.
Throughout in this paper we assume a gas consists of atomic hydrogen. 
For the range of temperatures considered, since the gas is almost neutral, we adopt $\kappa=2.5\times10^3\sqrt{T}$
cm$^{-1}$ K$^{-1}$ s$^{-1}$ \citep{P53}.
In this paper, we adopt the following fitting formula of the net cooling rate \citep{KI02},
\begin{equation}
        \rho{\cal L}(\rho,T) = \frac{\rho}{m_\mathrm{H}}\left( -\Gamma + \frac{\rho}{m_\mathrm{H}} \Lambda(T) \right)\;\;\mathrm{erg\;cm^{-3}\;s^{-1}},
 \label{cooling rate}
\end{equation}
\begin{equation}
        \Gamma = 2\times10^{-26}\;\;\mathrm{erg\;s^{-1}}, \nonumber
\end{equation}
\begin{equation}
        \frac{\Lambda(T)}{\Gamma} = 10^7\exp\left( -\frac{118400}{T+10^3} \right) 
        + 1.4\times10^{-2}\sqrt{T}\exp\left( -\frac{92}{T} \right) \nonumber.
\end{equation}
Figure \ref{fig:eq} shows 
the thermal equilibrium state of this net cooling rate in the $(n,P)$ plane. 
The gas is subject to the cooling (heating) above (below) the thermal equilibrium curve.
The bistable fluid consisting of the WNM and CNM 
can exist in the pressure range of $P_\mathrm{min}<P<P_\mathrm{max}$, where 
$P_\mathrm{min}/k_\mathrm{B}=1596$ K cm$^{-3}$ and $P_\mathrm{max}/k_\mathrm{B}=5012$ 
K cm$^{-3}$.
The saturate pressure where the static front is realized is as large
as $P_\mathrm{sat}/k_\mathrm{B}=2823$ K cm$^{-3}$ in the plane-parallel geometry.

In steady state $(\partial/\partial t=0)$, equations (\ref{eoc}) and (\ref{eom}) 
can be integrated with respect to $x$ to give 
\begin{equation}
        \rho v = j
        \label{const}
\end{equation}
and
\begin{equation}
        \rho v^2 + P = M,
        \label{const mom}
\end{equation}
respectively.
The energy equation (\ref{eoe}) becomes
\begin{equation}
        c_pj \frac{dT}{dx}
        -v \frac{dP}{dx} - \frac{d}{dx}\left( \kappa\frac{dT}{dx } \right)+\rho{\cal L}(\rho,T)=0,
        \label{integ eq0}
\end{equation}
where $j$ and $M$ denotes the mass and momentum fluxes, respectively, that are spatially constant, 
and $c_p=\gamma {\cal R}/(\gamma-1)$ is the specific heat at constant pressure.
From equations (\ref{eos}), (\ref{const}), and (\ref{const mom}), density, pressure, and velocity can be expressed by using 
temperature, the mass flux, and momentum flux as follows:
\begin{equation}
        \rho = \frac{1}{2{\cal R}T}\left( M + \sqrt{M^2 - 4 {\cal R}j^2 T } \right),
        \label{rho}
\end{equation}
\begin{equation}
        v = \frac{1}{2j}\left( M - \sqrt{M^2 - 4{\cal R} j^2 T}  \right),        
        \label{v}
\end{equation}
\begin{equation}
        P = \frac{1}{2}\left( M + \sqrt{M^2 - 4 {\cal R}j^2 T } \right).        
        \label{P}
\end{equation}
For convenience, instead of $T$, we introduce new variable $\theta$ defined by $d\theta=\kappa dT$.
Using $\theta$ and equations (\ref{rho})-(\ref{P}), equation (\ref{integ eq0}) can be rewritten as
\begin{equation}
         \frac{d^2\theta}{dx^2}  = \frac{ {\cal R}j}{2\kappa(\theta)}\left[ \frac{\gamma+1}{\gamma-1} 
        + \frac{M}{\sqrt{M^2-4{\cal R}j^2T(\theta)}}\right] \frac{d\theta}{dx}
        +\rho{\cal L}.
        \label{integ eq}
\end{equation}

\citet{SI01} and \citet{IIK06} described a detailed numerical method to solve equation (\ref{integ eq}).
Since equation (\ref{integ eq}) is the second order differential equation, we need 
two boundary conditions. At $x=-\infty$, we consider a uniform CNM with thermal equilibrium, 
${\cal L}(\rho_\mathrm{c},P_\mathrm{c})=0$, where physical variables in the CNM are denoted by 
the subscript of `c'. The boundary conditions are given by 
\begin{equation}
        \theta(x=-\infty)=\theta_\mathrm{c},\;\;\mathrm{and}\;\;\;\frac{d\theta}{dx}\Bigr|_{x=-\infty}=0.
        \label{bou}
\end{equation}
We have two free parameters: $P_\mathrm{c}$ and $j$. Given $P_\mathrm{c}$, the density and temperature 
can be obtained from equation (\ref{eos}) and the equilibrium condition 
${\cal L}(\rho_\mathrm{c},T_\mathrm{c})=0$.
The momentum flux is determined by $P_\mathrm{c}$ and $j$ from equations (\ref{const}) and (\ref{const mom}).

The properties of differential equation (\ref{integ eq}) can be well understood 
as trajectories in the phase diagram
$(\theta,d\theta/dx)$. From equation (\ref{integ eq}), one can see that the spatially uniform state with 
thermally equilibrium (${\cal L}=0$) corresponds to a stationary point because $d\theta/dx=d^2\theta/dx^2=0$.
The topological property of the CNM at $x=-\infty$ corresponds to saddle point \citep{Eetal92,FS93}.
Given $P_\mathrm{c}$ and $j$, we numerically integrate equation (\ref{integ eq}) from $x=-\infty$ along 
one of the eigenvectors obtained from the Jacobi matrix of equation (\ref{integ eq}) around the 
stationary point of the CNM.

\section{Zel'dovich \& Pikel'ner Solutions}\label{sec:ZP}
First, we review the solutions connecting two thermal equilibrium states (CNM and WNM). 
We call the solutions derived by ZP69 the ZP solutions.
The ZP solutions exist only in the 
pressure range of $P_\mathrm{min}<P_\mathrm{c}<P_\mathrm{max}$ (see figure \ref{fig:eq}).
Given $P_\mathrm{c}$, \citet{SI01} determine $j=j_\mathrm{ZP}(P_\mathrm{c})$ as the eigenvalue problem 
so that the boundary condition $d\theta/dx=0$ is satisfied in the WNM at $x=\infty$.

For the case of a static solution, that is $j=v=0$, equation (\ref{integ eq}) can be rewritten as
\begin{equation}
        \int_{\theta_\mathrm{c}}^{\theta_\mathrm{w}} \rho {\cal L}d\theta
        = \int_{T_\mathrm{c}}^{T_\mathrm{w}}\kappa\rho{\cal L}dT = 0
 \label{saturate}
\end{equation}
(ZP69)
where the subscript `w' denotes the physical quantities in the WNM.
Equation (\ref{saturate}) shows the balance between cooling and heating inside the transition layer.
The pressure satisfying equation (\ref{saturate}) is called saturation pressure.

In steady solutions with non-zero mass flux $j\ne0$, 
integrating equation (\ref{integ eq0}) from $x=-\infty$ to $x=\infty$, one obtains
\begin{equation}
        j = - \frac{q}{c_p \left( T_\mathrm{w} - T_\mathrm{c}\right)},\;\;\mathrm{where}\;\;
        q = \int_{-\infty}^{\infty} \rho{\cal L}(\rho,T)dx, 
\label{j}
\end{equation}
and we neglect the second term of equation (\ref{integ eq0}) since the flow is subsonic and hence 
almost isobaric. 
In the saturation pressure, $q=0$ since $j=0$.
For $P>P_\mathrm{sat}$, $q$ is positive because cooling dominates heating inside the front. 
Thus, $j<0$ from equation (\ref{j}), i.e. the front corresponds to the condensation.
In contrast, for $P<P_\mathrm{sat}$, the front describes the evaporation.
In the ZP solutions, the mass flux $j$ is very small. 
Thus, the energy is mainly transferred by heat conduction.
From equation (\ref{integ eq0}), one can see that the thickness of the front is characterized by 
the Field length \citep{BM90} defined by 
\begin{equation}
        l_\mathrm{F}\equiv \sqrt{\frac{\kappa T}{|\rho {\cal L}|}}.
\end{equation}
The typical values of $l_\mathrm{F}$ is as large as $10^{-3}$ pc in the CNM, and $10^{-1}$ pc in the WNM.
From equation (\ref{j}), the flow speed across the transition layer 
can be evaluated by $|v|\simeq l_\mathrm{F}/t_\mathrm{c}$, 
where $t_\mathrm{c}$ is the cooling timescale \citep{FS93}.
Since this value is very small in the actual ISM, ZP69 suggested that the transition front
is almost static.

Figure \ref{fig:j0}(a) represents properties of the ZP solution for 
$P_\mathrm{c}/k_\mathrm{B}=3000$ K cm$^{-3}$.
The corresponding mass flux is as large as $j_\mathrm{ZP}=-3.2\times10^{-3}m_\mathrm{H}$ km s$^{-1}
(v_\mathrm{w}=4.7\times10^{-3}$ km s$^{-1}$).
The first row of figure \ref{fig:j0} shows trajectories of solutions with various boundary conditions 
in the phase diagram $(\theta,d\theta/dx)$. There are three stationary points, which are indicated 
by the filled circles. The CNM and WNM are denoted by `C' and `W', respectively.
As mentioned in section \ref{sec:numerical}, the CNM and WNM correspond to saddle points.
On the other hand, the stationary point at the intermediate $\theta$ corresponds to the unstable phase denoted 
by `U' in figure \ref{fig:j0}(a), and it is spiral point \citep{Eetal92,FS93}.
The solutions starting from the CNM are denoted by the thick solid line. The ZP solution connects two
saddle points `C' and `W'.
The second row of figure \ref{fig:j0}(a) represents the temperature profile. 
The trajectory of the ZP solution in the $(n,P)$ plane is shown in the third row of figure \ref{fig:j0}(a).

\section{Dynamical Transition Layers}\label{sec:dynamical}
\subsection{Transition Layers for $P_\mathrm{min}<P_\mathrm{c}<P_\mathrm{max}$}\label{sec:zp large}
In the ZP solutions, the mass flux is determined so that the CNM is connected with the WNM.
Here, we consider the condensation front $(j<0)$. What happen if the mass flux is larger than
$|j_\mathrm{ZP}|$? 
\subsubsection{Classification of Solutions}
\begin{figure*}
        \begin{center}
                  \includegraphics[width=13.0cm]{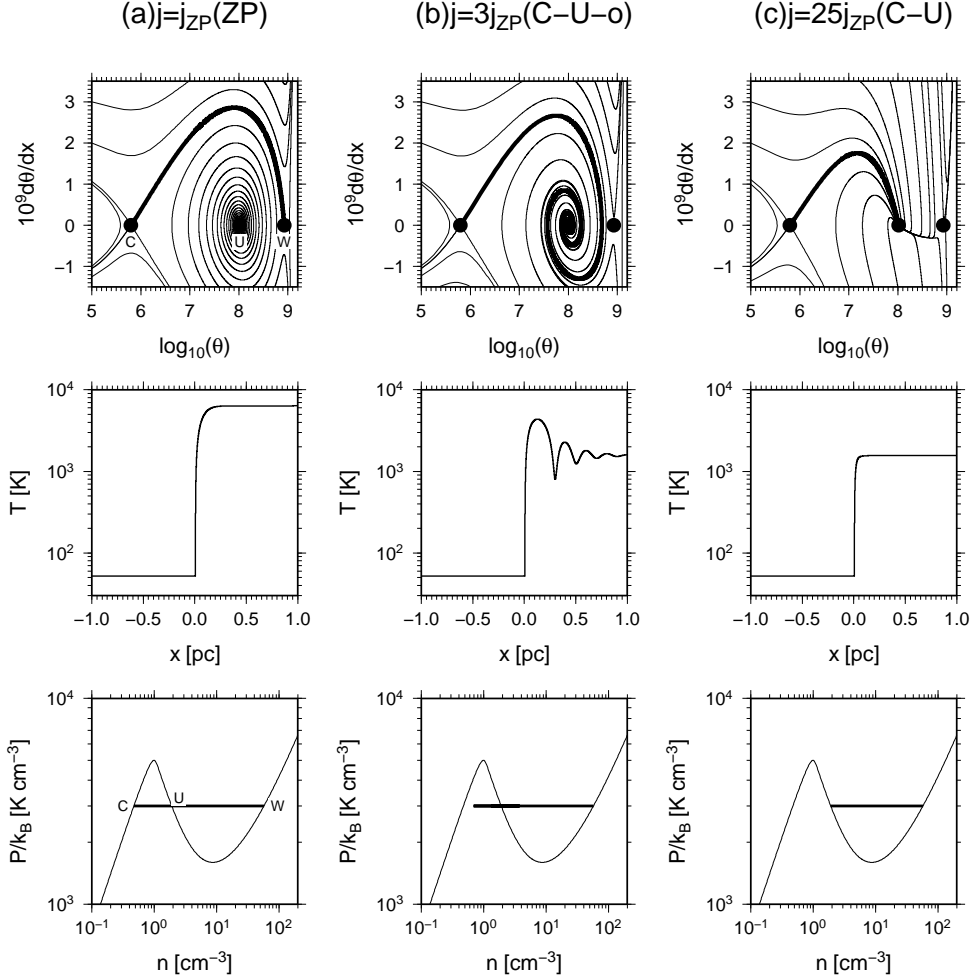}
        \end{center}
        \caption{
        Properties of solutions for $j/j_\mathrm{ZP}=1$(a), $3$(b), and $25$(c).
        The first row indicates the trajectories of solutions with various boundary conditions.
        The obtained solution satisfying equation (\ref{bou}) are shown by the thick lines.
        The labels `C', `U', and `W' denote the CNM, unstable phase, and WNM.
        The second row shows the temperature profile in each mass flux.
        The third row represents trajectory of the obtained solution in the $(n,P)$ plane.
        }
        \label{fig:j0}
\end{figure*}
It is found that solutions for larger mass flux can be divided by four type solutions: 
C-U-o, C-U, C-c, and C-C.
We describe the property of each solution below.

{\it C-U-o Solutions.}
For the case with $j=3j_\mathrm{ZP}$, the solution does not reaches the saddle point of the WNM but approaches
the spiral point of the unstable phase 
(see the first row of figure \ref{fig:j0}b). From the second row of figure \ref{fig:j0}(b), 
one can see the oscillation of the temperature profile that corresponds to the spiral motion in the 
phase diagram. The third row shows the trajectory of the solution in the $(n,P)$ plane.
The solution truncates at the heating-dominated region before arriving at the WNM.
The truncation point corresponds to the peak of the temperature in the second row of figure \ref{fig:j0}(b).
Since the mass flux is still low, the pressure is almost constant.
This type solutions already has been found in \citet{Eetal92} under the isobaric approximation and 
a more simplified cooling rate.
\citet{IIK06} also found them as the finite extent solutions,
and they truncated the solution at the first peak of the temperature profile near the transition front.
We call this type of solutions `C-U-o' solutions, where `C-U' means solutions connecting 
CNM and unstable phase, and `o' means the oscillation.

\begin{figure}
        \begin{center}
                  \includegraphics[width=8.0cm]{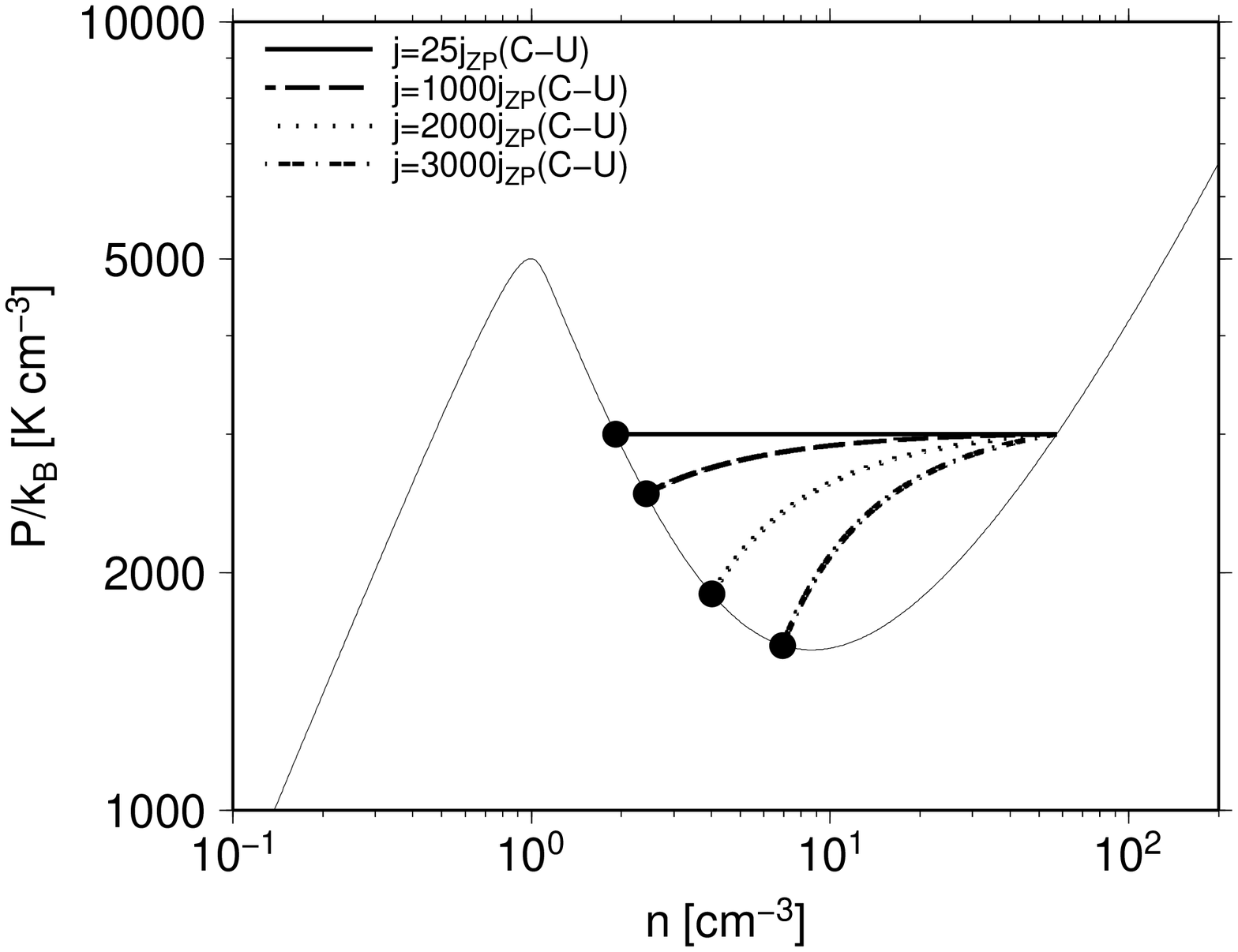}
        \end{center}
        \caption{
        Trajectories of solutions for $j/j_\mathrm{ZP}=1$(the solid line), 1000(the dashed line), 
        2000(the dotted line), 3000(the dot-dashed line). 
        The CNM pressure is set to $P_\mathrm{c}=3000k_\mathrm{B}$ K cm$^{-3}$.
        The filled circle shows 
        the state at $x=\infty$.
        }
        \label{fig:large_flux}
\end{figure}

{\it C-U Solutions.}
For larger mass flux $j=25j_\mathrm{ZP}$, the topological properties of the stationary point of the unstable phase
changes from spiral to node (see the first row of figure \ref{fig:j0}c)
\footnote[1]{
To be precise, before the C-U-o solution switches into the C-U solution, 
the spiral point already changes into the node point. 
In this paper, we categorize solutions with temperature peak as the C-U-o solution instead of 
the difference of the topological property of the unstable phase.
}.
Thus, from the second row of figure \ref{fig:j0}(c), 
there is no oscillation in the temperature profile,
and the CNM is monotonically connected with the unstable phase.
Figure \ref{fig:large_flux} shows solutions for much larger mass flux, $j/j_\mathrm{ZP}=25$, 1000, 2000, and 
3000. One can see that as the mass flux increases, the density (pressure) of the unstable phase at 
$x=\infty$ increases (decreases) along the equilibrium curve. Because of large mass flux,
the pressure increases significantly toward the CNM. 
This type solutions with lower mass flux already has been found in \citet{Eetal92} under
isobaric approximation.
We call this type of solutions `C-U' solutions.

\begin{figure}
        \begin{center}
                  \includegraphics[width=8.0cm]{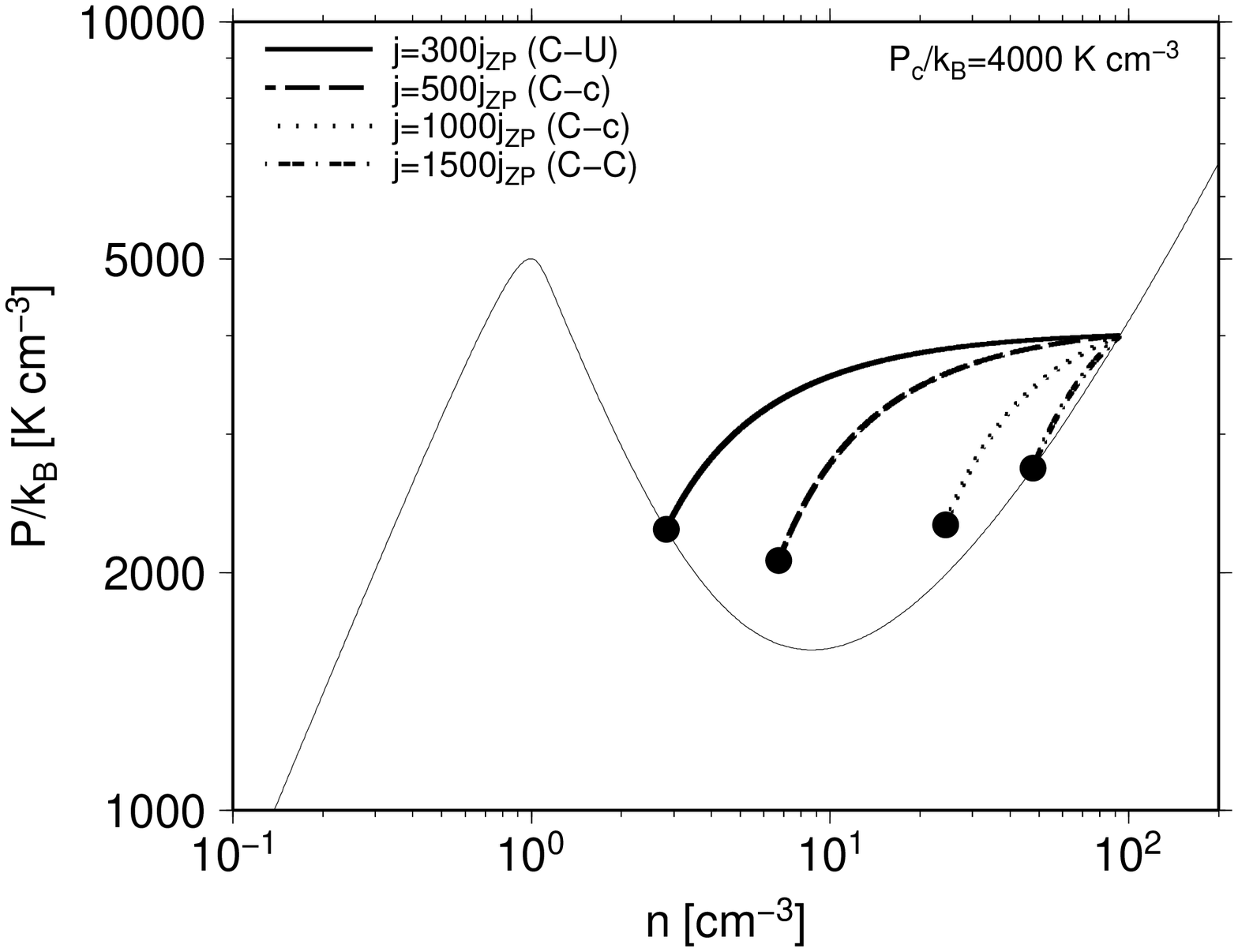}
        \end{center}
        \caption{
        Trajectories of solutions for $j/j_\mathrm{ZP}=300$(the solid line), 500(the dashed line), 
        1000(the dotted line), 1500(the dot-dashed line). 
        The CNM pressure is set to $P_\mathrm{c}=4000k_\mathrm{B}$ K cm$^{-3}$.
        The filled circle shows 
        the state at $x=\infty$.
        }
        \label{fig:large_flux_sonic}
\end{figure}

{\it C-c Solutions.}
Figure \ref{fig:large_flux_sonic} shows the solutions for $j/j_\mathrm{ZP}=300$, 500, 1000, and 1500.
Here, we set $P_\mathrm{c}/k_\mathrm{B}=4000$ K cm$^{-3}$.
The corresponding mass flux is $j_\mathrm{ZP}=-2.15\times10^{-2}m_\mathrm{H}$ km s$^{-1}$. The solution for $j=300j_\mathrm{ZP}$ belongs
to the C-U solution.
For larger mass flux $j=500j_\mathrm{ZP}$, the solution truncates at a certain point before arriving at the equilibrium state.
This critical point corresponds to $T=T_\mathrm{crit}=M^2/(4{\cal R}j^2)$ above which solutions do not
exist (see equation (\ref{rho})-(\ref{P})).
The physical variables at the critical point is denoted by using the subscript of `crit'.
From equations (\ref{const}) and (\ref{const mom}), the condition $T=T_\mathrm{crit}$ is equivalent 
to 
\begin{equation}
        |v_\mathrm{crit}| = \sqrt{P_\mathrm{crit}/\rho_\mathrm{crit}}.
        \label{vcrit}
\end{equation}
Note that equation (\ref{vcrit}) is slightly smaller than the adiabatic sound speed $\sqrt{\gamma P_\mathrm{crit}
/\rho_\mathrm{crit}}$. These critical velocity and temperature are also seen in the 
steady state structure of the shock front with heat conduction without physical viscosity 
\footnote{
The inclusion of physical viscosity may change the properties and existances of these solutions.
}\citep{ZR67}.
At the critical point, one can see that equation (\ref{integ eq}) becomes infinity. Thus, in order to 
obtain a finite solution, $d\theta/dx$ must vanish at the critical point.
From equations (\ref{P}) and (\ref{integ eq}), the pressure gradient becomes finite as follows:
\begin{eqnarray}
\left( \frac{dP}{dx} \right)_{\mathrm{crit}} &=& - \left( \frac{2{\cal R}j^2}{\sqrt{M^2-4{\cal R}j^2T}} 
        \frac{dT}{dx}\right)_\mathrm{crit} \\
        &=& - \left( \rho{\cal L}_\mathrm{crit} \right)\sqrt{\frac{\rho_\mathrm{crit}}{P_\mathrm{crit}}}\nonumber
        \label{}
\end{eqnarray}
The pressure at the critical point can be derived from the momentum flux conservation.
At the CNM $(x=-\infty)$, since the velocity is still subsonic because of its high density, the momentum flux 
becomes
\begin{equation}
        M=\rho_\mathrm{c}v_\mathrm{c}^2 + P_\mathrm{c}\simeq P_\mathrm{c}.
\end{equation}
On the other hand, from equation (\ref{P}), the pressure at the critical point is given by $P_\mathrm{crit}=M/2$.
Therefore, the pressure at the critical point can be expressed by $P_\mathrm{c}$ as follows:
\begin{equation}
        P_\mathrm{crit}\simeq P_\mathrm{c}/2.
        \label{P crit}
\end{equation}
The density at the critical point is given by
\begin{equation}
        \rho_\mathrm{crit}=j^2/P_\mathrm{crit}\simeq 2j^2/P_\mathrm{c}.
        \label{rho crit}
\end{equation}
One can see that $P_\mathrm{crit}$ for $j=500j_\mathrm{ZP}$ and 1000$j_\mathrm{ZP}$ 
is roughly equal to $P_\mathrm{c}/2=2000k_\mathrm{B}$
K cm$^{-3}$, and its density increases for larger mass flux as seen in equation (\ref{rho crit}).
Around the critical point, since $dT/dx=d\theta/dx/\kappa=0$, the trajectories in the $(n,P)$ plane 
approach constant temperature lines of $T=T_\mathrm{crit}$ in figure \ref{fig:large_flux_sonic}.

We investigate the topological properties at the critical point in the phase diagram.
The Jacobi matrix of equation (\ref{integ eq}) around the critical point becomes infinite.
Thus, there is no solution passing the critical point smoothly.
In order to obtain infinite extent solution, shock front is required.
Solutions with shock front are described in Appendix \ref{app:shock}.
We call this type of solutions ``C-c'' solution, where ``C-c'' means solutions connecting 
the CNM and the critical point.

{\it C-C Solutions.}
For much larger mass flux case $j=1500j_\mathrm{ZP}$, since the solution passes the equilibrium curve before 
arriving at the critical point, the solution becomes the infinite extent solution connecting 
the CNM and CNM (see figure \ref{fig:large_flux_sonic}).
We call this type of solutions C-C solutions, where `C-C' means solutions connecting the CNM and CNM.

\begin{figure}
        \begin{center}
                  \includegraphics[width=8.0cm]{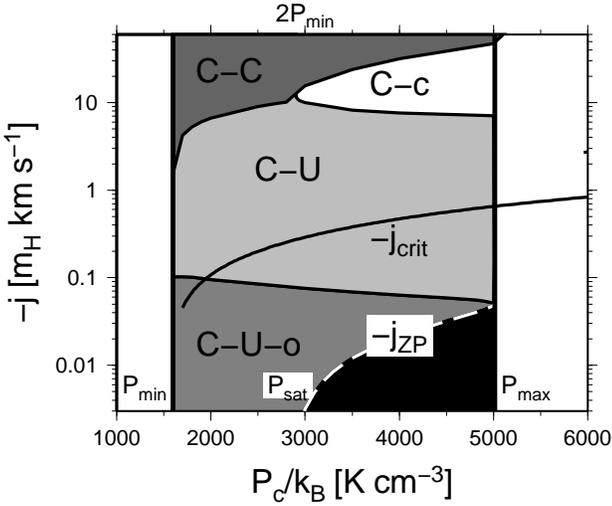}
        \end{center}
        \caption{
        Classification diagram of obtained solutions in parameter space $(P_\mathrm{c},j)$.
        The white dashed line corresponds to $j_\mathrm{ZP}$.
        The solid line indicates the critical mass flux $j_\mathrm{crit}(P_\mathrm{c})$.
        The black area where $|j|<|j_\mathrm{ZP}|$ is not focused in this paper.
        }
        \label{fig:sol_diag_lowpre}
\end{figure}
Figure \ref{fig:sol_diag_lowpre} shows classification of solutions in the parameter space $(P_\mathrm{c},j)$
for $P_\mathrm{min}<P_\mathrm{c}<P_\mathrm{max}$.
The white dashed line corresponds to $j=j_\mathrm{ZP}(P_\mathrm{c})$ on which the ZP solutions exist.
The black area where $|j|<|j_\mathrm{ZP}|$ is not focused in this paper. From figure \ref{fig:sol_diag_lowpre}, one can 
see that C-c solutions exist only for $P_\mathrm{c}>2P_\mathrm{min}$. This is because
if $P_\mathrm{c}<2P_\mathrm{min}$, the solutions meet the equilibrium state before arriving 
at the critical point (see equation (\ref{P crit})).

\subsubsection{Structure of Transition Layers}\label{sec:jcrit}
We investigate how the structure of transition layers depends on the mass flux.
As mentioned in section \ref{sec:ZP}, in the ZP solutions, the thickness of the front is
characterized by the Field length because the mass flux is very small.
As $|j|$ increases, the effect of advection (the first term of equation (\ref{integ eq0}))
becomes important compared with the heat conduction (the third term of equation (\ref{integ eq0}))
in the energy transfer. We can define a critical mass flux $j_\mathrm{crit}$ where 
the effect of advection becomes comparable to the heat conduction as follows:
\begin{equation}
        - c_pj_\mathrm{crit}\frac{T_\mathrm{m}}{l_\mathrm{F,m}} = \kappa \frac{T_\mathrm{m}}{l_\mathrm{F,m}^2}
        \Rightarrow j_\mathrm{crit}=-\frac{l_\mathrm{F,m}}{\gamma l_\mathrm{c,m}} \rho_\mathrm{m}c_\mathrm{m},
        \label{jcrit}
\end{equation}
where the subscript `m' denotes a reference state in the solutions, 
$l_\mathrm{c,m}$ is the cooling length defined by 
\begin{equation}
        l_\mathrm{c,m} = c_m \frac{P_\mathrm{m}}{(\gamma-1)\rho_\mathrm{m}{\cal L}_\mathrm{m}},
\end{equation}
and $c_\mathrm{m}=\sqrt{\gamma P_\mathrm{m}/\rho_\mathrm{m}}$ is the sound speed.
As the reference state, we consider the state where $|\cal L|$ has the maximum value for $T>T_\mathrm{c}$ under
the constant pressure $P=P_\mathrm{c}$. This state roughly gives minimum Field length and cooling length 
in the solution.
Thus, the critical mass flux is a function of $P_\mathrm{c}$. The actual value of $j_\mathrm{crit}$ is shown by 
the solid line in figure \ref{fig:sol_diag}. In most range of $P_\mathrm{c}$, $j_\mathrm{crit}$ lies in
the region occupied by the C-U solutions.

\begin{figure}
        \begin{center}
                  \includegraphics[width=8.0cm]{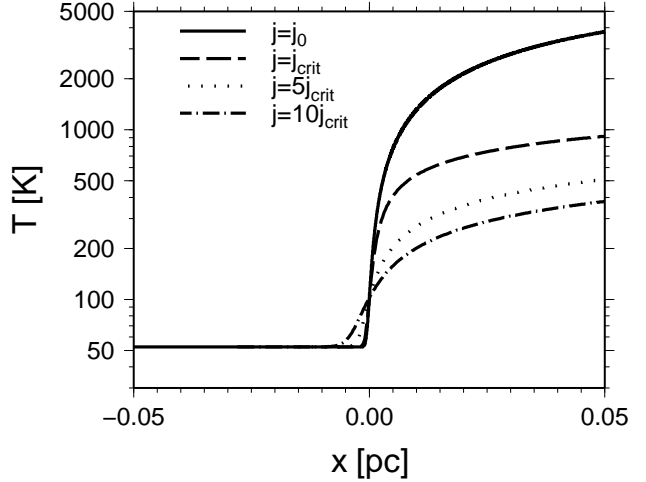}
        \end{center}
        \caption{
        Temperature profiles for $j=j_\mathrm{ZP}$(the solid line), $j_\mathrm{crit}$(the dashed line), 
        $5j_\mathrm{crit}$(the dotted line), and 
$10j_\mathrm{crit}$(the dot-dashed line). 
The CNM pressure is set to $P_\mathrm{c}/k_\mathrm{B}=300$ K cm$^{-3}$.
        }
        \label{fig:jcrit_profile}
\end{figure}

For $|j|>|j_\mathrm{crit}|$, advection dominates heat conduction in energy transfer.
In this case, the thickness of the front is characterized by the cooling length instead of the Field length.
In typical ISM, $l_\mathrm{F}/l_\mathrm{c}$ is roughly as small as $10^{-2}$.
Therefore, as the mass flux increases, the thickness of the front increases.
Figure \ref{fig:jcrit_profile} shows the temperature profiles for $j=j_\mathrm{ZP}$, $j_\mathrm{crit}$, $5j_\mathrm{crit}$, and 
$10j_\mathrm{crit}$. The CNM pressure is set to $P_\mathrm{c}/k_\mathrm{B}=3000$ K cm$^{-3}$.
One can see that the thickness of the layer increases with the mass flux.

\subsection{Transition Layers for $P_\mathrm{c}>P_\mathrm{max}$}\label{sec:gt Pmax}
In previous section, we found solutions only for $P_\mathrm{min}<P_\mathrm{c}<P_\mathrm{max}$.
However, in actual ISM, the CNM pressure can be greater than $P_\mathrm{max}$, for example,
in shock-compressed regions. Our solutions can be directly extend for $P_\mathrm{c}>P_\mathrm{max}$.

\begin{figure}
        \begin{center}
                  \includegraphics[width=8.0cm]{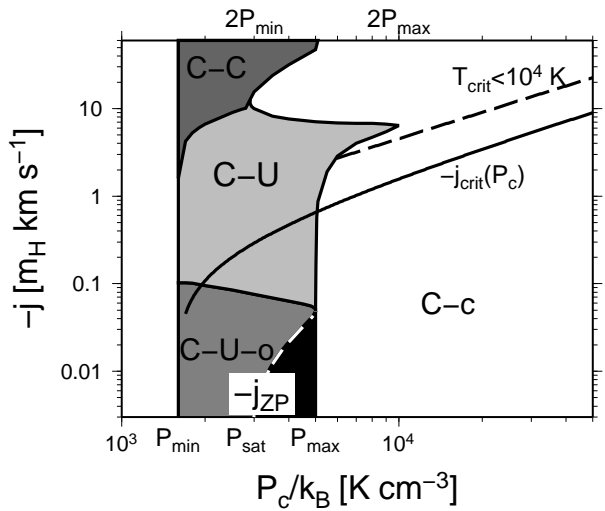}
        \end{center}
        \caption{
The same as figure 
\ref{fig:sol_diag_lowpre} but including the pressure range of $P_\mathrm{c}>P_\mathrm{max}$.
The dashed line corresponds to $T_\mathrm{crit}=10^4$ K.
        }
        \label{fig:sol_diag}
\end{figure}

\begin{figure}
        \begin{center}
                  \includegraphics[width=8.0cm]{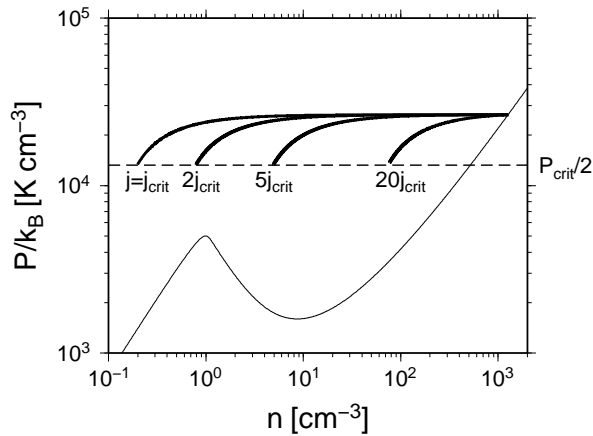}
        \end{center}
        \caption{
        Trajectories of solutions for $j/j_\mathrm{crit}=1$, 2, 5, and 20.
        The dashed line corresponds to $P=P_\mathrm{crit}/2$.
        }
        \label{fig:large_flux_hp}
\end{figure}

Figure \ref{fig:sol_diag} is the same as figure 
\ref{fig:sol_diag_lowpre} but it includes the pressure range of $P_\mathrm{c}>P_\mathrm{max}$.
Figure \ref{fig:sol_diag} shows that most of the regions are 
covered by the C-c solution for $P_\mathrm{c}>P_\mathrm{max}$.
The area of C-U solutions spreads to the pressure range of $P_\mathrm{max}<P_\mathrm{c}<2P_\mathrm{max}$.
This is because solutions in this pressure range can connect with thermal equilibrium states before 
reaching at the critical point at a certain range of $j$.
Figure \ref{fig:large_flux_hp} shows trajectories of the solutions for 
$j/j_\mathrm{crit}=1$, 2, 5, and 20. The CNM pressure is set to $2.65\times10^{4}$ K cm$^{-3}$. 
One can see that the solution terminates at the critical point
whose pressure is equal to $P_\mathrm{c}/2$ and where density increases with the mass flux.

\section{Discussion}\label{sec:discuss}
\subsection{Implications}
\citet{KI06} have investigated multi-dimensional nonlinear development of the TI without 
external forcing. They found that the turbulence is self-sustained by the TI.
The velocity dispersion for sufficiently large simulation box is as large as $\sim0.2-0.4$ km s$^{-1}$.
They suggested that this velocity is much larger than the prediction in ZP69.
Our solutions exist in such a large mass flux.
From their simulation, the mass flux becomes 
$|j|\sim 0.2\;m_\mathrm{H}\;\mathrm{km\;s^{-1}}(n/0.5\;\mathrm{cm^{-3}})(v/0.4\;\mathrm{km\;s^{-1}})$,
where we assume that the accretion velocity to the CNM is comparable to this velocity dispersion.
From figure \ref{fig:sol_diag_lowpre}, the corresponding solutions lie in the C-U or C-U-o solutions.

\citet{KI02} have investigated the TI in the shock-compressed region. 
They found that the velocity dispersion is as large as several km s$^{-1}$.
The corresponding mass flux is 
$|j|\sim 20\;m_\mathrm{H}\;\mathrm{km\;s^{-1}}(n/10\;\mathrm{cm^{-3}})(v/2\;\mathrm{km\;s^{-1}})$ that 
is larger than the results in \citet{KI06}.
The pressure of the shock compressed region is larger than $P_\mathrm{max}$, indicting that the two stable phases
cannot coexist. 
In shock compressed regions, the CNM clouds are embedded not by the stable WNM but by the unstable gas that
is supplied continuously by shock compression.
The C-c solutions in this pressure range $(P>P_\mathrm{max})$ 
may describe dynamical condensation from the unstable gas to the CNM in shock compression regions.


\subsection{Curvature Effect}
Our solutions are derived by assuming the plane-parallel geometry.
\citet{Netal05} investigated the curvature effect of the transition front by assuming 
the quasi-steady approximation.
They consider the cylindrical and spherical CNM cloud at the centre.
In contrast to the plane parallel solutions, 
the solutions have three parameters, the pressure of the CNM, the velocity of the front with respect 
to the centre, and the cloud radius. 
The second parameter corresponds to the mass flux in our solutions.
They solved equations as the eigen- and boundary-value problem so that the solutions connect the CNM and WNM.
Thus, the velocity of the front $v_f(P_\mathrm{c},R_\mathrm{c})$ 
is a function of $P_\mathrm{c}$ and cloud radius $R_\mathrm{c}$.
The curvature effect enhance the heat flux by the thermal conduction from the CNM to the WNM.
Thus, smaller cloud that has larger curvature tends to evaporate quickly.
On the other hand, cold gas that has a negative curvature tends to gain mass by the condensation.
Thus, the evolution of CNM cloud strongly depends on its shape.

By using the quasi-steady approximation, we can easily take into account curvature effect in our solutions.
It is expected that solutions with larger evaporation rate connect the unstable phase and WNM while 
solutions with larger condensation rate connect the CNM and unstable phase.


\section{Summary}\label{sec:summary}
In this paper, we have investigated steady condensation solutions of phase transition layers 
in large parameter space 
$(P_\mathrm{c},j)$ much larger than previous works.
We summarise our results as follows:

\begin{itemize}
        \item In the pressure range where the three thermal equilibrium phase can coexist under 
              a constant pressure $(P_\mathrm{min}<P_\mathrm{c}<P_\mathrm{max})$,  
              we find solutions that connects the CNM and the unstable phase.
              The solutions can be classified into four type solutions, C-U-o, C-U, C-c, and C-C solutions in
              order of the mass flux.
              The C-U-o and C-U solutions connect the CNM and the unstable phase with and without 
              the oscillation of physical quantities in the low-density side.
              The C-c solution is truncated at the maximum temperature above which there are no steady 
              solutions. Combination of a C-c solution and a shock front provides an infinite 
              extent solution.  The C-C solutions connect the CNM and CNM.

       \item We have also found steady solutions 
               in the pressure range $(P_\mathrm{c}>P_\mathrm{max})$
               where only CNM can be in equilibrium.
              This pressure range is realized in the shock compressed region. 
              In this parameter space $(P_\mathrm{c},j)$, most of solutions belong to the C-c solution.

       \item We derived a critical mass flux $j_\mathrm{crit}$ above which the effect of advection becomes
             important compared with the heat conduction in energy transfer.
             For large mass flux, $|j|>|j_\mathrm{crit}|$, the thickness of the front is characterized by 
             the cooling length instead of the Field length, so that the thickness becomes larger.
\end{itemize}

\section*{Acknowledgement}
We thank the referee for valuable comments.
We thank Dr. Tsuyoshi Inoue and Dr. Jennifer M. Stone for variable discussions.
This work was supported by Grants-in-Aid for Scientific Research from the MEXT of Japan
(K.I.:22864006; S.I.:18540238 and 16077202).


\appendix
\section{Solutions with Shock Front}\label{app:shock}
In this Appendix, we present steady solutions with shock front.
Here, we assume that preshock gas is in thermal equilibrium state.
The physical variables in preshock gas are denoted by using the subscript of ``E''.
The position of the shock front $x=x_\mathrm{sh}$ is determined so that 
the Rankine-Hugoniot relation is satisfied between the 
solutions and the thermal equilibrium state by the following method.
Given $x=x_\mathrm{sh}$, 
the specific energy flux is given by 
\begin{equation}
\frac{\gamma P_\mathrm{sh}}{(\gamma-1)\rho_\mathrm{sh} } + \frac{1}{2}v_\mathrm{sh}^2 \equiv \frac{\gamma+1}{2(\gamma-1)}
c_*^2,
\end{equation}
where the subscript of ``sh'' denotes physical variables at $x=x_\mathrm{sh}$, and 
$c_*^2$ is the effective sound speed.
The simple relation among $v_\mathrm{sh}$, $v_\mathrm{E}$, and $c_*$ is 
given by 
\begin{equation}
        v_\mathrm{sh}v_\mathrm{E}=c_*^2.
        \label{v}
\end{equation}
From equation (\ref{v}), one can get $v_\mathrm{E}$, and the Mach number ${\cal M}$ is obtained.
From the Rankine-Hugoniot relations, the physical variables in the preshock gas
($\rho_\mathrm{E}$ and $T_\mathrm{E}$) are obtained.
In general, the preshock gas is not in thermal equilibrium state.
The position of the shock front $x_\mathrm{sh}$ is determined iteratively until 
${\cal L}(\rho_\mathrm{E},T_\mathrm{E})=0$
is satisfied by using the bisection method.

\begin{figure*}
        \begin{center}
                  \includegraphics[width=10.0cm]{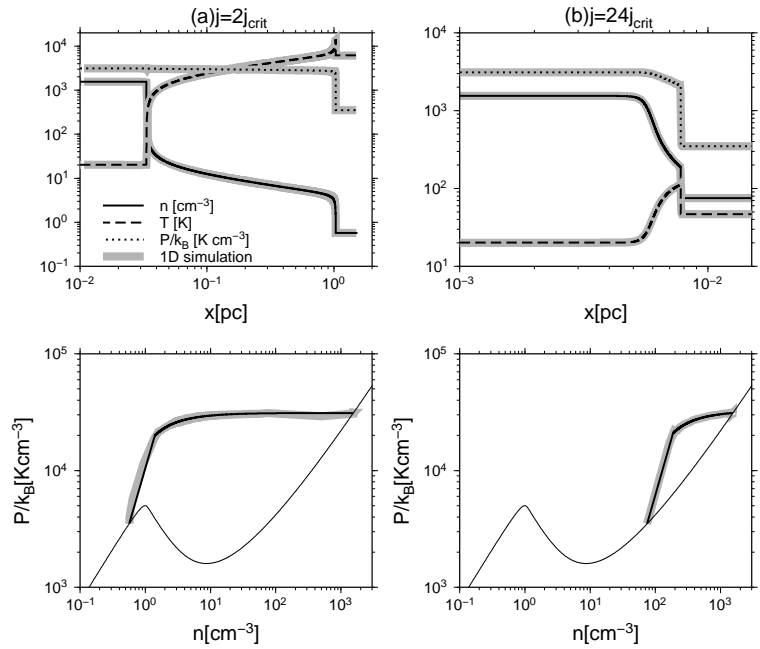}
        \end{center}
        \caption{
        {\it Upper panels:} 
        Density (solid line), temperature (dashed line), and pressure (dotted line) profiles 
        of steady state solutions at which $P_\mathrm{c}=3.113\times10^4$ K cm$^{-3}$ 
        for (a)$j=2j_\mathrm{crit}$ and (b)$j=24j_\mathrm{crit}$.
        {\it Lower panels:} Trajectories of steady state solutions in the $(n,P)$ plane.
        The thick gray line in each panel shows the results of one-dimensional hydrodynamical 
        simulation.
        }
        \label{fig:mesh}
\end{figure*}

As examples, we consider two cases of $j=2$ and 24$j_\mathrm{crit}$ 
for $P_\mathrm{c}/k_\mathrm{B}=3.114\times10^4$K cm$^{-3}$.
For $j=2j_\mathrm{c}$ case, the corresponding physical quantities of the preshock gas
are $n_\mathrm{E}=0.57$ cm$^{-3}$ 
and ${\cal M}=2.17$. 
For $j=24j_\mathrm{c}$, they  are $n_\mathrm{E}=75$ cm$^{-3}$ and ${\cal M}=2.23$.
Thus, the preshock gases belong to the WNM for $j=2j_\mathrm{c}$ and CNM for $j=24j_\mathrm{crit}$.
The upper panels of figures \ref{fig:mesh}a and \ref{fig:mesh}b show the 
density, temperature, and pressure distribution for $j=2j_\mathrm{crit}$ and $j=24j_\mathrm{crit}$, 
respectively. The lower panels of figure \ref{fig:mesh} show the trajectories of obtained 
solutions in the $(n,P)$ plane. 

In addition, we perform one-dimensional hydrodynamical simulation in order to confirm the steady
state solutions with shock front are realized.
We consider colliding wall problem where the gas flows collide at $x=0$
with $|v_\mathrm{E}-v_\mathrm{c}|$ from $x=\pm\infty$.
Two shock wave propagates from $x=0$ outward. Shock heated gas quickly cools until the gas reaches 
the thermal equilibrium state. After that, the cold layer forms and its thickness grows by gas accretion.
The thick gray line in each panel of figure \ref{fig:mesh} 
indicate the results of one-dimensional simulations in the growing phase 
of the cold layer.
One can see that the results of one-dimensional simulations 
are well described by the steady-state solutions.

\end{document}